\documentclass[11pt]{article}
\usepackage{moriond,epsfig}

\def\Journal#1#2#3#4{{#1} {\bf #2}, #3 (#4)}

\def\HPA{\em Helv. Phys. Acta}
\def\GRG{\em Gen. Rel. Grav.}
\def\NCA{\em Nuovo Cimento}

\def\NPB{{\em Nucl. Phys.} B}
\def\PLB{{\em Phys. Lett.}  B}
\def\PRL{\em Phys. Rev. Lett.}
\def\PRD{{\em Phys. Rev.} D}
\def\ZPC{{\em Z. Phys.} C}
\def\EPJC{{\em Eur. Phys. J.} C}

\def\gev{\, \mbox{GeV}\,}
\def\KL{K\"all\'en-Lehmann\,}

\def\be{\begin{equation}}
\def\ee{\end{equation}}
\def\bea{\begin{eqnarray}}
\def\eea{\end{eqnarray}}

\begin{document}
\vspace*{4cm}
\title{NEW SPECTRA IN THE HEIDI MODELS}

\author{ J.J. VAN DER BIJ }

\address{Institut f\"ur Physik, Albert-Ludwigs Universit\"at Freiburg, H. Herderstr. 3,\\
79104 Freiburg i.B., Deutschland}

\maketitle\abstracts{ We study the so-called HEIDI models, which are renormalizable extensions of the
standard model with a higher-dimensional scalar singlet field. We compare their predictions
with the recent results from the LHC. We show that the data can easily be described
by the HEIDI models. However more data are necessary in order to distinguish these models
from the standard model. A particular prediction is that the width of the Higgs could be
in the GeV range. Such a width could be difficult to establish at the LHC. 
Suggestions for experiments beyond the
LHC are made.}

\section{Introduction}
The standard model is in very good agreement with the data.
With the exception of the Higgs sector all particles of the model have been seen
and sofar there is no evidence for new physics. There appear to be no new 
fermions, vectorbosons or new forces. The results from the experiments at the
LHC, presented in this conference, have put strong constraints on the possibility 
of new physics. As there are strong theoretical arguments, that the standard model is
the only possible low-energy theory for the vectorbosons and the fermions \cite{cosmo1,cosmo2,cosmo3},
this is not surprising.

 This leaves only the Higgs-sector for a possible discovery.
In order to determine what one should look for, let us discuss the situation before
the presentation on 13 december 2011 at CERN. What was known for sure is, that massive
vectorbosons exist with non-abelian couplings. For quantum mechanics to be valid this implies
that a Higgs field has to exist \cite{quigg1,quigg2}, otherwise one violates unitarity.
Since quantum field theory is correct, this implies that the Higgs field must satisfy the
\KL representation \cite{kal,leh}, so it has a spectral density. Because of renormalizability the integral over
the spectral density must be one. Without this condition the precision electroweak data
cannot be  described properly. Actually the precision electroweak data show that the 
spectral density is concentrated in the low mass region; the Higgs boson is light. 
This much is basically  certain about the Higgs sector. Everything else is conjecture.
In particular the idea, that there is a single Higgs particle delta-peak in the spectrum 
is an assumption, for which there is no basis in theory or experiment. As the Higgs field
appears in the theory in a different way from the other fields, it is not unreasonable
to assume that the Higgs field has a non-trivial spectral density. It is somewhat ironic that
this possibility was overlooked \cite{englert} in the 1960's, when the Higgs particle was introduced,
as in this period there was a lot of research on the related subject of dispersion relations.
The scientific goal regarding the electroweak symmetry breaking is therefore to measure
the \KL spectral density of the Higgs field. Under circumstances this might be quite difficult
at the LHC and one would have to think about a Higgs-factory beyond the LHC.
In the following we introduce the HEIDI models, which form an elegant construction 
allowing for non-trivial spectral densities with a minimum number of parameters.
Then we compare the models with the new results from the LHC and discuss future prospects.

\section{Lagrangian and propagator}
The fact that propagators in quantum field theory are described by a \KL spectral density
follows from fundamental principles of quantum mechanics and relativity and is mathematically
 rigorously true. Normally however
one ignores this, since most fields correspond directly to single particles and
one takes a Breit-Wigner approximation for the propagator. The general idea of a spectral
density is somewhat abstract. In order to clarify the idea and make contact with more
usual descriptions I will below give Lagrangians, where such a \KL representation arises naturally.
First we start with a discrete spectrum and then generalize to a (partly) continuous spectrum. 
The basic ideas have been presented in Moriond before \cite{moriond2007,moriond2008}.

\subsection{The Hill model}
The simplest extension of the standard model, with only two extra parameters is the
Hill \cite{hill} model. One adds to the standard model a single scalar field $H$ with the
following Lagrangian.
\begin{equation}
{\cal L}  = -\frac{1}{2} (D_\mu \Phi)^{\dagger}(D_\mu \Phi) - \lambda_1/8
(\Phi^\dagger \Phi -f_1^2)^2
 { -\frac{1}{2}(\partial_\mu H)^2 - \frac{\lambda_2}{8} (2 f_2 H} - \Phi^\dagger \Phi)^2.
\label{eq:hill}
\end{equation}
Here $\Phi$ is the standard model Higgs doublet.
Important for the structure of the theory is that the extra singlet has no self-couplings.
Somewhat surprisingly this is consistent with renormalizability.
After diagonalizing the Lagrangian one ends up with two particles similar to the
standard model Higgs particle, however with  
the couplings to the standard model particles reduced by a common factor,
so the branching ratios do not change. An alternative way to describe these features
is to use a modified  
propagator for the Higgs field, containing two poles.
\begin{equation}
D_{HH}(k^2) = \frac {sin^2 \alpha}{k^2 + m_+^2} + \frac {cos^2 \alpha}{k^2 + m_-^2}
\label{eq:hillprop}
\end{equation}
Such a description is sufficient to study Higgs signals (interaction basis).

The generalization to more fields is straightforward.
One introduces 
n Higgses ${H_i}$ with reduced couplings $g_i$ to the standard model particles.
After diagonalizing one finds \cite{akhoury,gunion} the sum rule:
\begin{equation}
\Sigma g_i^2 = g^2_{Standard~model}
\label{eq:sumrule}
\end{equation}
This can straightforwardly be generalized to a continuum density $\rho (s)$,
with the relation:
\begin{equation}
\int  \rho (s) ds  = 1
\label{eq:KLintegral}
\end{equation}
The density  $\rho (s)$ is the 
K\"all\'en Lehmann density, which has here been constructed from a tree level Lagrangian.

\subsection{HEIDI models}
Given enough fields one can construct an arbitrary spectral density for the
Higgs field, however this involves  an infinite number of parameters as well,
which is not very satisfactory. The question is whether there is a more elegant way.
Indeed there is. Because the singlet fields have no self-couplings all their
interactons are superrenormalizable in four dimensions. Therefore one 
can take a higher dimensional field for the Hill field without 
spoiling renormalizability. This explains the name HEIDI 
(german for high-D, compare with SUSY for supersymmetry).
One can go up to six-dimensional fields without spoiling renormalizability and
can even postulate fields with a fractional dimension. 
With this procedure one still has only a few parameters,
but can get a fairly broad range of spectra.
For deriving the propagator one assumes a torus compactification giving
 rise top winding modes $H_i$;  however we will take the continuum limit in the end.

In terms of the modes {$H_i$} the Lagrangian is the following:

\begin{equation}
\begin{array}{rcl}
L&=&{-\frac{1}{2}D_\mu \Phi^\dagger D_\mu \Phi-\frac{M_0^2}{4}\Phi^\dagger \Phi -\frac{\lambda}{8}(\Phi^\dagger \Phi)^2}\\
&-&\frac{1}{2}\sum (\partial_\mu H_k)^2-\sum \frac{m_k^2}{2} H_k^2\nonumber\\
&-& {\frac{g}{2} \Phi^\dagger \Phi} \sum H_k-\frac{\zeta}{2}\sum H_i H_j ~~~
\end{array}\label{eq:modelagrangian}
\end{equation}

 $m_k^2=m^2+ m_{\gamma}^2\,\vec k^2$, where $\vec k$ is a $\gamma$-dimensional
vector, $m_{\gamma}=2\pi/L$ and $m$  a $d$-dimensional mass term for the field $H$.

In terms of continuum fields the last two terms can be described by the following action:

\begin{equation}
 S=\int{d^{4+{\gamma}}x} \prod_{i=1}^{\gamma} \delta(x_{4+i})\left(g_B H(x) 
{\Phi^{\dagger}\Phi} -\zeta_B H(x) H(x)\right)  
\label{eq:continuumaction}
 \end{equation}  

\subsection{Propagators}
Minimizing the potential, diagonalizing and taking the continuum limit,
one finds the following propagator:

\begin{equation}
D_{HH}(q^2)=\Bigg(q^2 +M^2-\frac{\mu^{8-d}}{(q^2+m^2)^{\frac{6-d}{2}} \pm \nu^{6-d}} \Bigg)^{-1}
\label{eq:propagatornd}
\end{equation}
 
Here $M$ is a fourdimensional mass, $m$ a higher dimensional mass, $\mu$ a high-to-fourdimensional
mixing scale and $\nu$ a brane mass, mixing the higher-dimensional fields among themselves.
 This term is new in the analysis and is needed, because it is
induced by renormalization in any case.

To go to six dimensions one has to make a limiting procedure and finds a propagator in the
following form:

\begin{equation}
D_{HH}(q^2)=\Bigg(q^2+M^2 + \mu^{2} \frac{\log((q^2+m^2)/m^2)}{1 + \alpha_6\,\log((q^2+m^2)/m^2)}\Bigg)^{-1}
\label{eq:propagator6d}
\end{equation}

Given the form of the propagator it is straightforward to find the spectral density, by taking the imaginary
part. There are some constraints on the parameters, while tachyons must be absent. 
Dependent on the parameters there are different possibilities. One can have zero, one or two
delta-peaks, corresponding to "particles". At a mass greater than the location of the peak(s)
there is a continuum,
that starts at $s=m^2$.

\section{Comparison with experiment}
As reported in previous Moriond conferences \cite{moriond2007,moriond2008}
it is possible to compare the HEIDI models with Higgs search experiments.
We first discuss the situation before the LHC and then see how
the new data change the description compared to the previous results.

\subsection{Before the LHC}
Before the LHC data the HEIDI models were compared with the direct
Higgs search at LEP-200.
Within the pure standard model the absence of a clear signal led to
a lower limit on the Higgs boson mass of $114.4 \gev$ at the 95\% confidence level.
Although no clear signal was found the data had some intriguing features,
that could be interpreted as evidence for Higgs bosons beyond the standard
model. There is a $2.3\,\sigma$ effect seen by all experiments at around $98 \gev$.
A somewhat less significant $1.7\,\sigma$ excess was seen around $115 \gev$. Finally
over the whole range $s^{1/2} > 100\gev$ the confidence level was less than
expected from background.
These features were taken as evidence for a spread-out Higgs-boson \cite{bijlep,pulice}.
The peak at $98 \gev$ was taken to correspond to the delta peak in the
\KL density. The other excess data were interpreted as part of the continuum,
that peaks around $115 \gev$. Fitting the data with this interpretation led to the picture
that the Higgs signal would be a very broad enhancement at low energies, that could
escape detection at the LHC.

\subsection{Interpretation of the LHC data}
With the results for the Higgs search at the LHC the picture has changed.
The excess at 115\gev, that was present at LEP-200 has now disappeared;
averaging between LEP, CMS and ATLAS one is very close to pure background.
However the somewhat more significant peak at 98\gev has not been affected by
the LHC data, which are not sensitive in this range yet. Actually the ATLAS
experiment even sees an excess in the 101\gev bin, which might be related to the
LEP peak. Of course this is not statistically very relevant, however we take it as a motivation
to keep the LEP peak at 98\gev in the analysis.
This leaves us with the excess in the 116-130\gev range.
The interpretation of the data is not quite straightforward.
Making a naive average of the CMS and ATLAS measurements one finds 
a picture that can be interpreted as a single peak, two peaks 
or a continuous signal with a bit of fluctuations. Clearly more data are necessary.
We will leave the question of the precise form of the signal open and interpret the combined 
LEP and LHC data in the following way. There is no signal below 95\gev,
there is a $2.3\,\sigma$ signal at 98\gev, there is no further signal below
116\gev and the bulk of the spectrum is between 116 and 130\gev.
Allowing for uncertainties in the experiment we therefore impose the following
conditions on the spectral density:
\begin{equation}
 95\gev < m_{peak} < 101\gev,~~~~~~~~~~~~~~~~
0.056 < g^2_{98}/g^2_{SM} < 0.144
\label{eq:lepdata}
\end{equation}
\begin{equation}
m > 116 GeV,~~~~~~~~~~~~~~~~~
 ~~~~~~~~~~\int_{(130)^2}^{\infty} \rho (s) ds < 0.1
\label{eq:lhcdata}
\end{equation}

\subsection{Fits to the data}
We first attempt to fit the data with one peak around 98\gev plus a continuum.
The results are shown in figures 1-4. The inner lines correspond to 
keeping a 10\% peak at 98\gev. For the outer lines we vary the location and strength
of the peak. From the figures it is clear that one can fit the data easily without 
any particular fine-tuning of parameters. 

How would the signal for a HEIDI Higgs differ from
a standard model Higgs? First there is of course the small peak around 98\gev. 
Another interesting effect is, that the continuum appears as a wide Higgs, with a  somewhat weaker
coupling than for 
a full Higgs particle peak. A typical spectrum is presented in figure 5 for a 
five-dimensional and a six-dimensional model. One sees that  the width of the Higgs is much larger
than in the standard model, it is roughly 2\gev for the parameters in the figure. 
In principle one would like to use the
data to distinguish between the two curves in the figure, but this might be difficult because of the experimental
resolution at the LHC.

One can also fit the data with two peaks and a continuum, one peak at 98\gev
and the second in the range 116-130\gev. In the experiment one can probaly not separate the second peak from the
continuum, as the continuum starts very close to the second peak. The situation is described
in figure 6, where we plot the start of the continuum as a function of the strength of the second
peak, together with the mass average over the combination of the second peak and the continuum.
One would probably see a somewhat wider and asymmetric peak only.

Finally it is possible to have a pure continuum, but in this case one has to ignore
the LEP peak, assuming that it is a statistical fluctuation.

\section{Future prospects}
It is clear that the present data are not enough to come to definite conclusions.
Below we discuss some questions to be addressed in the near and far future

\subsection{Questions for the LHC}
Even with the planned running of the LHC this year it is unlikely that
one can distinguish the HEIDI models from the standard model in a definite way.
What is clear from the above discussion is that the first priority should be
to clearly establish the presence of a signal, which might be a bit weaker 
than a standard model Higgs signal. The present analysis should be extended
to lower energies, in order to get to the LEP peak at 98\gev.
Combined data from ATLAS and CMS are highly desirable from the
theoretical point of view.
It would probably be possible to already start with a "model-independent"
analysis for the spectral density $\rho (s)$. A somewhat crude first approach
could be for instance to divide the range of 116-130\gev in 7 bins
of 2\gev each and allow for a Higgs signal to be present in the
bins for instance in steps of $1/6$ of a Higgs signal. This would give rise
to 1716 models, for each of which one can calculate a probability how well it fits
the data.

After that, determining the branching ratios with some precision is important
to confirm that one is dealing with a Higgs boson, that is similar to the standard model.
However it is particularly important to determine the width of the Higgs boson,
or more general its spectral density. The width should be studied in correlation
with the strength of the Higgs signal, as also invisible decay could lead to a wide Higgs\cite{binoth}.
However in this case the branching ratio to standard model particles is suppressed. 
The invisible decay particles could be candidates for dark matter \cite{pospelov}. 
These latter two  issues can probably be addressed with some accuracy
only when the LHC reaches its full design parameters. Even then the LHC will not be able to determine
the line shape of the Higgs boson to great accuracy, due to the resolution of the detectors.

\subsection{A Higgs factory}
Ultimately to determine the \KL density of the Higgs field precisely, the LHC is not
an optimal machine. One will need a Higgs factory. There are essentially three options,
all of them with some disadvantages. The first would be a muon-collider, where the Higgs 
is produced directly and one can make an energy scan to probe the resonance region
as in LEP with the Z-boson.
 The disadvantage is that it is quite unclear
if such a machine can be built, in particular it would be difficult to get a high enough luminosity.
The second possibility would be a 250-300\gev linear collider. The problem here is that one needs a high
precision on the momenta of incoming and outgoing particles, as the spectrum has to be
determined from the recoil in the process $e^+e^-\rightarrow ZH$. Beamstrahlung would probably
reduce the resolution to an unacceptable level. However this should be studied in more detail.
Most studies for a linear collider are focused on 500\gev. It would be useful to study an optimal design
for Higgs physics. 

The problem of beamstrahlung can be largely avoided by going to a very large circular collider
with a CM energy of 250-300\gev. Synchrotron radiation drives up the
radius of the ring and
the largest collider proposed sofar is the very large lepton collider at Fermilab, which could have
a 230 km circumference. The limiting factor could be the precision with which one can measure
the momentum of the outgoing leptons coming from the Z-boson. This is a detector problem, that also
plays a role for the linear collider option, and needs careful study.
 
Although the civil engineering cost of such a large tunnel will be high, 
from the accelerator point of view a circular electron-positron collider 
is the easiest to build and could be achieved with current technology. 
In the longer term, a facility of this size would make a very natural 
stepping stone to study proton-proton collisions at very high energies, 
far beyond any LHC upgrade. A facility of this type would provide
an exciting physics program for O(40) years and would require 
cooperation at an unprecedented 
level on the global scale. Given the long time scales involved, it would 
be interesting to explore the physics case for a $\sqrt{s} \sim 
250-300$~GeV electron positron collider to fully explore a Higgs signal 
seen at the LHC, followed with a $50-100$~TeV proton-proton collider to 
probe the high energy frontier.

\section{Conclusion}
The conclusion can be short: Higher dimensions may be hidden in the
Higgs lineshape.\\
 Or somewhat more poetically:
\vskip0.2cm
\begin{center}
{\it \large Where~is~Heidi~hiding~?}\\
\vskip 0.1cm
{\it \large Heidi~is~hidden}\\
\vskip 0.1cm
{\it \large in~the~high-D~Higgs~Hill~!}\\
\end{center}

\section*{Acknowledgments}
This work was supported by the BMBF within the Verbundprojekt
HEP-THEORIE.

\newpage

\begin{figure}
\begin{center}
\psfig{figure=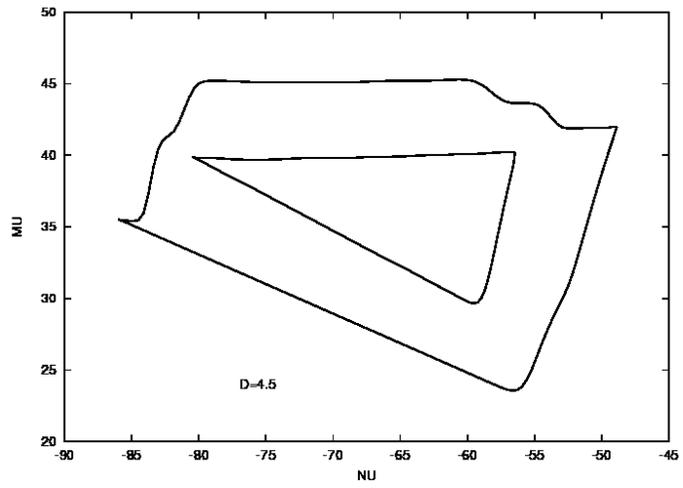,height=6.5cm}
\end{center}
\caption{Fit in 4.5 dimensions.
\label{fig:cont45}}
\end{figure}

\begin{figure}
\begin{center}
\psfig{figure=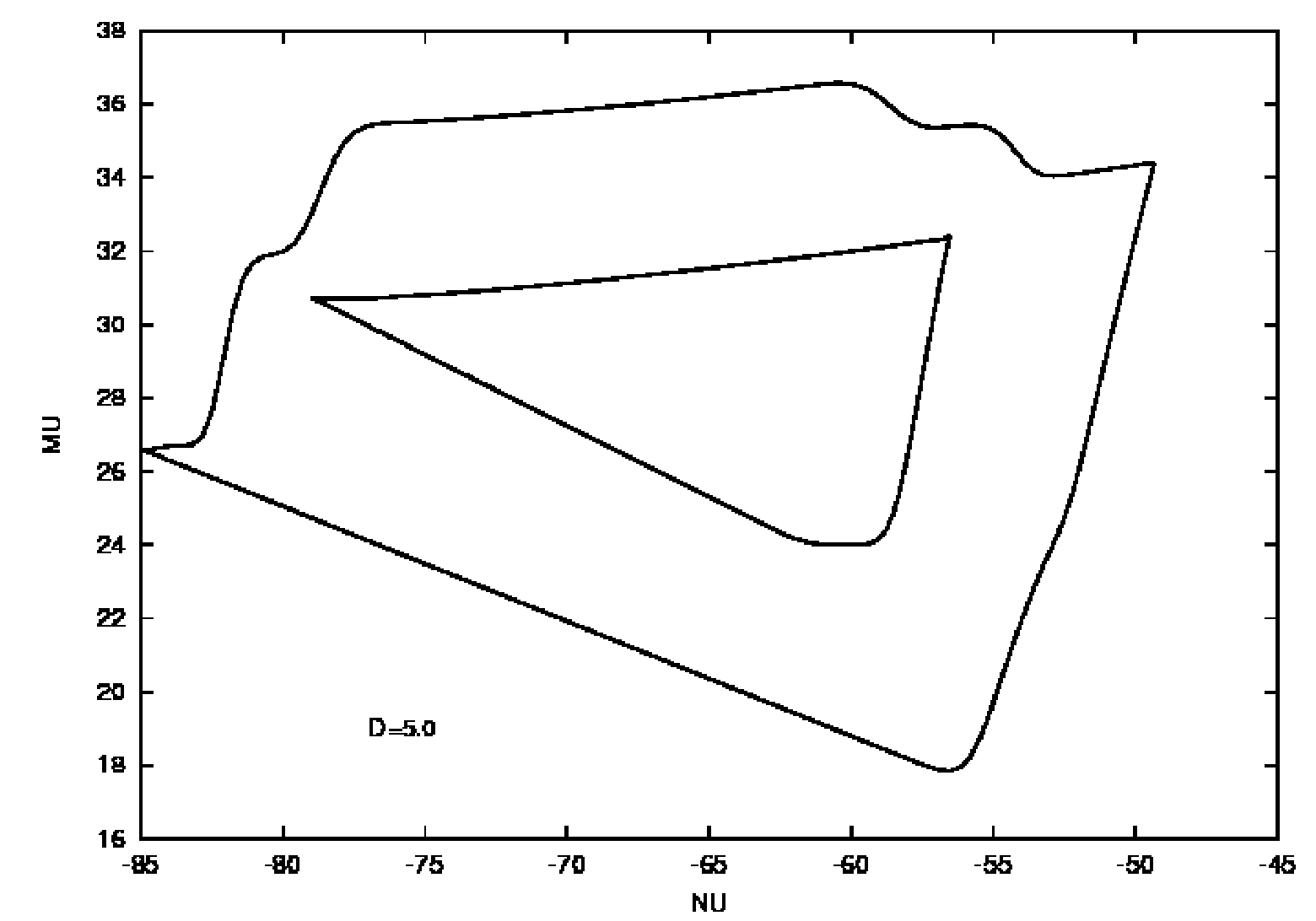,height=6.5cm}
\end{center}
\caption{Fit in 5.0 dimensions.
\label{fig:cont50}}
\end{figure}

\begin{figure}
\begin{center}
\psfig{figure=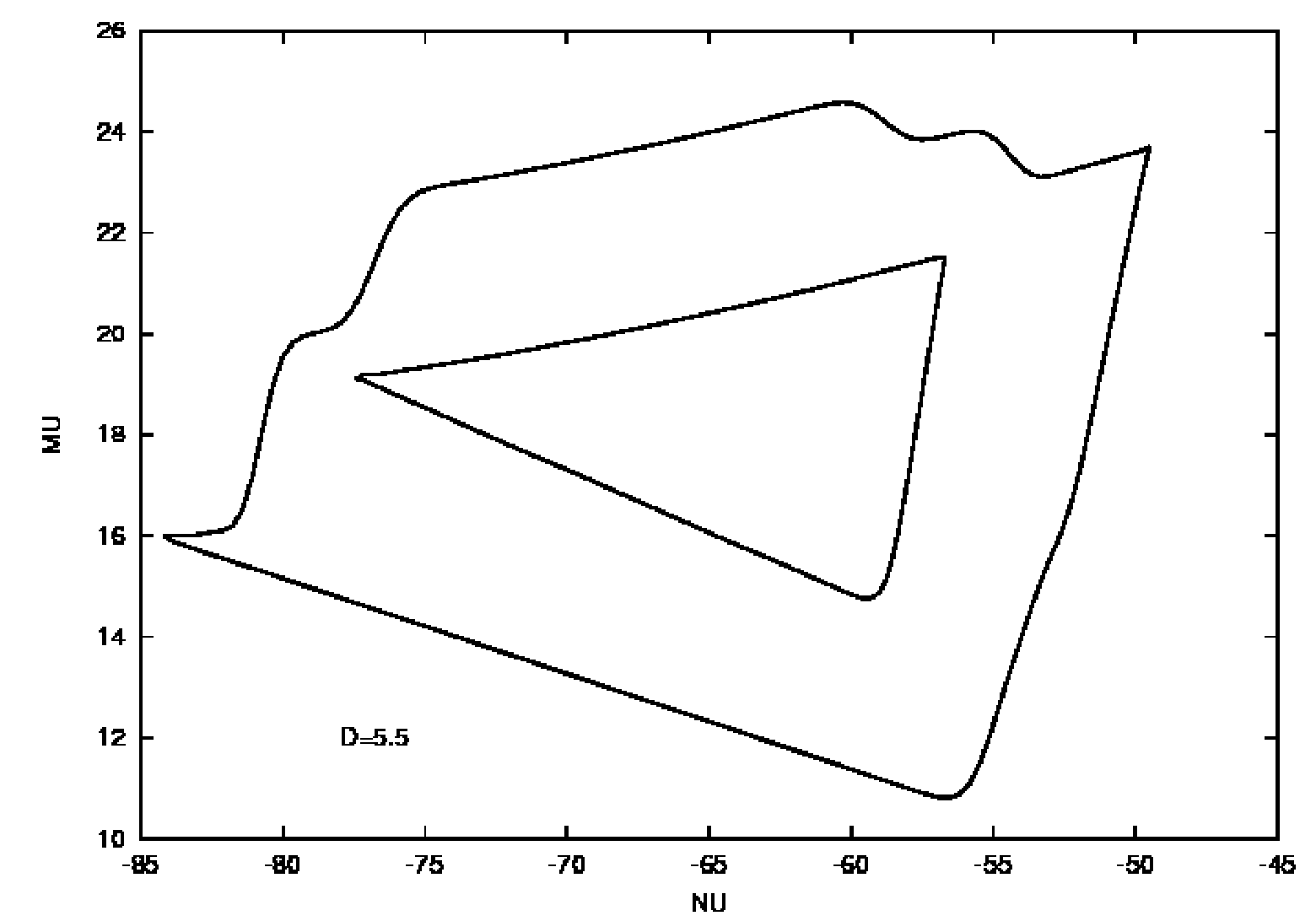,height=6.5cm}
\end{center}
\caption{Fit in 5.5 dimensions.
\label{fig:cont55}}
\end{figure}

\begin{figure}
\begin{center}
\psfig{figure=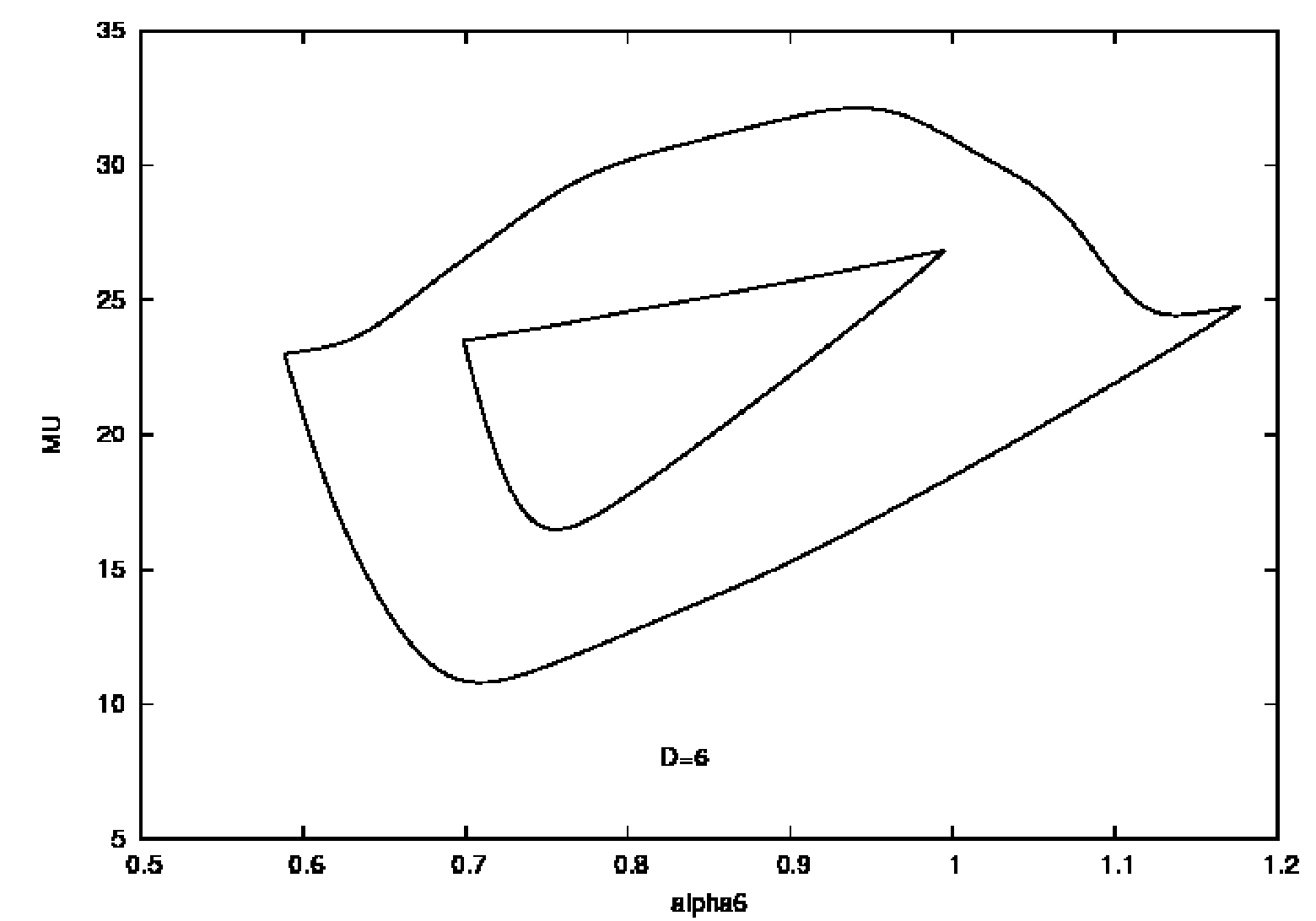,height=6.5cm}
\end{center}
\caption{Fit in 6.0 dimensions.
\label{fig:cont60}}
\end{figure}

\begin{figure}
\begin{center}
\psfig{figure=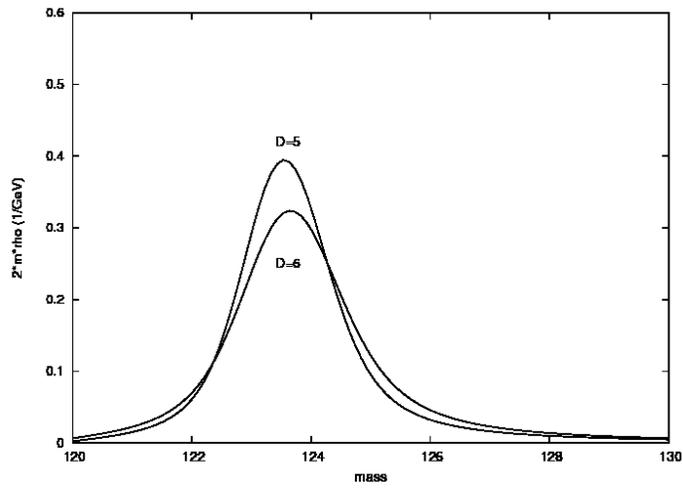,height=6.5cm}
\end{center}
\caption{Higgs line shape.
\label{fig:spectr56}}
\end{figure}

\begin{figure}
\begin{center}
\psfig{figure=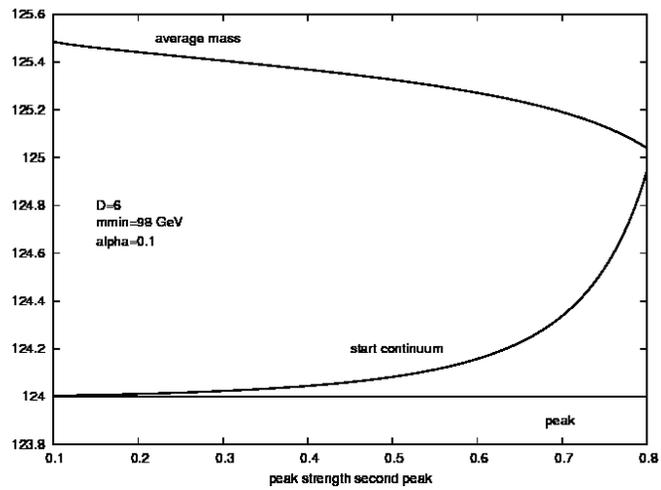,height=6.5cm}
\end{center}
\caption{Two peak structure.
\label{fig:twopeak}}
\end{figure}

\end{document}